%% file: lrpaper-fp113.tex
\documentclass{sig-alternate-2013}
\input{macros}
\usepackage{balance}

\def\sharedaffiliation{\end{tabular}\newline\begin{tabular}{c}}
\def\upf{\superscript{*}}
\def\yhoo{\superscript{\dag}}
\def\mi{\superscript{\S}}

\newfont{\mycrnotice}{ptmr8t at 7pt}
\newfont{\myconfname}{ptmri8t at 7pt}
\let\crnotice\mycrnotice%
\let\confname\myconfname%

\permission{Permission to make digital or hard copies of all or part of this work for personal or classroom use is granted without fee provided that copies are not made or distributed for profit or commercial advantage and that copies bear this notice and the full citation on the first page. Copyrights for components of this work owned by others than ACM must be honored. Abstracting with credit is permitted. To copy otherwise, or republish, to post on servers or to redistribute to lists, requires prior specific permission and/or a fee. Request permissions from Permissions@acm.org.}
\conferenceinfo{SIGIR'15,}{August 09 - 13, 2015, Santiago, Chile.}
\copyrightetc{\copyright~2015 ACM. ISBN \the\acmcopyr}
\crdata{978-1-4503-3621-5/15/08\ ...\$15.00.\\
DOI: http://dx.doi.org/10.1145/2766462.2767704}

\begin{document}

\title{Local Ranking Problem on the BrowseGraph}


\numberofauthors{4}
\author{
   \alignauthor \hspace{-0.5cm} Michele Trevisiol\upf\yhoo\\
   \email{\hspace{-0.5cm} trevisiol@acm.org} 
   \alignauthor \hspace{-2cm}Luca Maria Aiello\yhoo\\
   \email{\hspace{-2cm} alucca@yahoo-inc.com}
   \alignauthor \hspace{-4cm} Paolo Boldi\mi\\
   \email{\hspace{-4cm} boldi@di.unimi.it}
   \alignauthor \hspace{-6cm}Roi Blanco\yhoo\\
   \email{\hspace{-6cm} roi@yahoo-inc.com}
   \sharedaffiliation
  \begin{tabular}{ccc}
	\affaddr{{\upf}Web Research Group}	& \affaddr{{\yhoo}Yahoo Labs} &	\affaddr{{\mi}Dip. di Scienze dell'Informazione} \\
	\affaddr{Universitat Pompeu Fabra} & \affaddr{London, UK} &	\affaddr{Univ. degli Studi di Milano}\\
	\affaddr{Barcelona, Spain}  &	\affaddr{}				& \affaddr{Milano, Italy} \\
  \end{tabular}
}
\maketitle

\begin{abstract}
The ``Local Ranking Problem'' (LRP) is related to the computation of a centrality-like rank on a \emph{local} graph, where the scores of the nodes could significantly differ from the ones computed on the \emph{global} graph. Previous work has studied LRP on the hyperlink graph but never on the \bgraph, namely a graph where nodes are webpages and edges are browsing transitions. 
Recently, this graph has received more and more attention in many different tasks such as ranking, prediction and recommendation. However, a web-server has only the browsing traffic performed on its pages (\emph{local} \bgraph) and, as a consequence, the local computation can lead to estimation errors, which hinders the increasing number of applications in the state of the art.
Also, although the divergence between the local and global ranks has been measured, the possibility of \textit{estimating} such divergence using only local knowledge has been mainly overlooked. These aspects are of great interest for online service providers who want to: {\it (i)} gauge their ability to correctly assess the importance of their resources only based on their local knowledge, and {\it (ii)} take into account real user browsing fluxes that better capture the actual user interest than the static hyperlink network. We study the LRP problem on a \bgraph\ from a large news provider, considering as subgraphs the aggregations of browsing traces of users coming from different domains. We show that the distance between rankings can be accurately predicted based only on structural information of the local graph, being able to achieve an average rank correlation as high as $0.8$.
\end{abstract}


\section{Introduction} \label{sec:introduction} 
The ability to identify the online resources that are perceived as important by the users of a website is crucial for online service providers. Metrics to estimate the importance of the page from the structure of online links between them are widely used: algorithms that compute the \textit{centrality} of the nodes in a network, such as PageRank~\cite{Page1998}, HITS~\cite{Kleinberg1999} and SALSA~\cite{Lempel2001}, have been employed extensively in the last two decades in a vast variety of applications. Born and spread in conjunction with the growth of the Web, they can determine a value of importance of a page from the complex network of links that surrounds it.
More recently, centrality metrics have been applied to \emph{browsing graphs}, (also referred to as \bgraphs\ \cite{Liu2008a,trevisiol12image,TrevisiolRecSys14}) where nodes are webpages and edges represent the transitions made by the users who navigate the links between them. Differently from the hyperlink networks, this data source provides the analyst a way of studying directly the dynamics of the navigational patterns of users who consume online content. Also, unlike hyperlinks, browsing traces account for the variation of consumption patterns in time, for instance in the case of online news where articles tend to become rapidly stale. Comparative studies have shown that centrality-based algorithms applied over \bgraphs\ provide higher-quality rankings compared to standard hyperlink graphs~\cite{Liu2009, Liu2008a}.

Most centrality measures aim at estimating the importance of a node, using information coming from the \textit{global} knowledge of the graph topology. Potentially the addition of new nodes and edges, can have a cascade effect on the centrality values of all other nodes in the network.
This fact entails high computational and storage cost for big net\hyp{}works. More critically, there are some situations in which a global computation on the entire graph is unfeasible, for example when the information about the entire network is unavailable or if only an estimation for specific web pages is required.
This is an important limitation in many real-world scenarios, where the graphs at hand are often very large (Web scale) and, most importantly, their topology is not fully known. This practical issue raises the problem of how well one can estimate the actual centrality value of a node by knowing only a local portion of the graph. This is known as the \textit{Local Ranking Problem} (LRP)~\cite{Chen2004}.
One of the questions behind LRP is whether it is possible to estimate efficiently the PageRank score of a web page using only a small subgraph of the entire Web~\cite{Bressan2011}. In other words, if one starts from a small graph around a page of interest and extends it with external nodes and arcs (\ie, those belonging to the whole graph), how fast will one observe the computed scores converging to the real values of PageRank?
We extend this line of work in the context of browsing graphs. For the first time we study the LRP on the \bgraph\ and shed some light on the bias that PageRank incurs {\it (i)} when estimating the centrality score of nodes in a \bgraph, and {\it (ii)} when only partial information about the graph is available. To achieve that, we monitor the browsing traffic of the news portal and we extract different browsing subgraphs induced by the browsing traces of users coming from different \textit{domains}, such as search engines (\eg, Google, Yahoo, Bing) and social networks (\eg, Facebook, Twitter, Reddit). In this setting, the local \bgraphs\ are the subgraphs induced by the different domains, and the global \bgraph\ is the one built using indistinctly all the navigation logs of the news portal.
We describe and evaluate models that tell apart a subgraph from the others just by looking at the behavior of a random surfer that navigates through their links.
The results show how it is possible to recognize the graph using only the very first few nodes visited by the users, because the graphs are very different among them (even if they are extracted from the same big log of the news portal).
The implication of this experiment is two-fold: first it highlights how navigation patterns of the users differ among these subgraphs. 
Second, we learn that it is possible to infer the user domain of origin from the very first browsing steps. This capability enables several types of services, including user profiling~\cite{chiarandini12discovering}, web site optimization~\cite{WeischedelWOW2006}, user engagement estimation~\cite{lehmann2013measuring}, and cold-start recommendation~\cite{TrevisiolRecSys14}, even when the referrer URL is not available (\eg when the user comes from mobile social media applications or URL shortening services).
Once we show that the subgraphs are different enough, we proceed to perform more involved experiments that we call ``Growing Rings''. We examine the behavior of the PageRank computed on the local and the global graphs. 
In order to study how the local PageRank converges to the global one, we apply some strategies of incremental addition (``growing'') of external nodes to these subgraphs (``rings''). 
Finally, we build on these findings by setting up a prediction experiment that, for the first time, tackles the task of estimating the reliability of the PageRank computed locally. We measure \textit{how much} the local PageRank diverges from the global one using only structural features of the local graph, usually available to the local service provider.
To sum up, the main contributions of this work are the following:
\begin{itemize}
  \item We study the $LRP$ on a large-scale \bgraph\ built from a very popular news website. To the best of our knowledge we are the first to tackle this problem on the increasingly popular \bgraph~\cite{TrevisiolRecSys14,trevisiol12image,chiarandini12discovering,Liu2008a}. We present an analysis of the convergence of the PageRank on the local graph to the global one, by incrementally expanding the local graph in a snowingreen fashion.
  \item We tackle the problem of discovering the referrer domain of a user session, when this information is missing or hidden. We show that this is possible using a random surfer model, which is able to tell the referrer domain with high accuracy, just after the very first browsing transitions. 
  \item We show that an accurate estimation of the distance between the local and global PageRank can be obtained looking at the structural properties of the local graph, such as degree distribution or assortativity. 
\end{itemize}

The remainder of the paper is organized as follows. In $\S$\ref{sec:relatedwork} we overview relevant prior work in the area and in $\S$\ref{sec:dataset} we describe our dataset and the extraction of the browsing graphs. In $\S$\ref{sec:analysis} we analyze the (sub-)graphs and we highlight their differences. In $\S$\ref{sec:lrp-analysis} we study the LRP problem on the \bgraph\ and compare the approximation accuracy of different graph expansion strategies. 
In $\S$\ref{sec:prediction} we present the prediction experiment of the PageRank errors of the local graph. Last, in $\S$\ref{sec:conclusions} we wrap up and highlight possible extensions to the work.

\section{Related Work} \label{sec:relatedwork} 
This work encompasses two main different research areas that we introduce shortly. Our focus is the {\it Local Ranking Problem} but our contribution relates also to previous work on browsing log data, especially the ones that investigate or make use of centrality-based algorithms.

\subsection*{Local Ranking Problem}
The \textit{Local Ranking Problem} (LRP) was first introduced by Chen \etal~\cite{Chen2004} in $2004$, who addressed the problem to approximate/update the PageRank of individual nodes, without performing a large-scale computation on the entire graph.
They proposed an approach that can tackle this problem by including a moderate
number of nodes in the local neighborhood of the original nodes. Furthermore,
Davis and Dhillon~\cite{DavisKDD06} estimated the global PageRank values of a
local network using a method that scales linearly with the size of the local
domain. Their goal was to rank  webpages in order to optimize their crawling
order, something similar to what was done by Cho \etal~\cite{Cho1998} who
instead selected the top-ranked pages first. However, this latter strategy results to be in contrast with Boldi \etal~\cite{BoldiWAW2004}, as they found that crawling first the pages with highest global PageRank actually perform worse, if the purpose is fast convergence to the real (global) rank values. In this work, we partial expand the local graph with the neighboring nodes with highest (local) PageRank showing an initial improvement on the convergence speed.
In $2008$ the problem was reconsidered by Bar-Yossef and Mashiach~\cite{Bar-Yossef2008}, where they simplified the problem calculating a local {\it Reverse PageRank} proving that it is more feasible and computationally cheaper, as the reverse natural graphs tend to have low in-degree maintaining a fast PageRank convergence.
Bressan and Pretto~\cite{Bressan2011} proved that, in the general case, an efficient local ranking algorithm does not exist, and in order to compute a {\it correct} ranking it is necessary to visit at least a number of nodes linear in the size of the input graph. They also raised some of the research questions tackled in our paper that we discuss in Section~\ref{sec:deltaprediction}.
They reinforce their findings in later work~\cite{BressanWWW13}, where they summarized two key factors necessary for efficient local PageRank computations: {\it exploring the graph non-locally} and {\it accepting a small probability error}. These two constraints are also considered in this paper in order to perform our experiments on the browsing graphs. 
When one wants to estimate PageRank in a local graph, the problem of
the missing information is tackled in various ways.
In~\cite{Bar-Yossef2008,Bressan2011} for example, the authors make use of a model called {\it link server} (also known as {\it remote connectivity server}~\cite{Bharat1998}), which responds to any query about a given node with all the in-coming and out-going edges and relative nodes. This approach, with the knowledge about the LRP, allows to estimate the PageRank ranking, or even the score, with the relative costs.
A similar problem was studied by Andersen \etal~\cite{Andersen2007}, where their
goal was to compute the PageRank contributions in a local graph motivated by the
problem of detecting link-spam: given a page, its PageRank contributors are the
pages that contribute most to its rank; contributors are used for spam
detection since you can quickly identify the set of pages that contribute significantly to the PageRank of a suspicious page.

The problem we consider here is different and largely
unexplored, because we are studying the PageRank of the different subgraphs based
on user browsing patterns.

\subsection*{BrowseGraph}
In recent years a large number of studies of user browsing traces have been
conducted. Specifically, in the last years there was a surge of interest in the
\bgraph, a graph where the nodes are web pages and the edges
represent the transitions from one page to another made by the navigation of the
users. Characterizing the browsing behavior of users is a valuable source of
information for a number of different tasks, ranging from understanding how
people's search behaviors differ~\cite{White07investigatingbehavioral}, ranking
webpages through search trails~\cite{Agichtein:2006:IWS:1148170.1148177,
White:2010:ASR:1835449.1835548} or recommending content items using past
history~\cite{Tsagkias:2012:LIM:2348283.2348330}.
A comparison between the standard hyperlink graph, based on the structure of
the network, with the browse graph built by the users' navigation
patterns, has been made by Liu \etal~\cite{Liu2008a, Liu2009}. 
They compared centrality-based algorithms
like PageRank~\cite{Page1998}, TrustRank~\cite{Gyongyi2004}, and
BrowseRank~\cite{Liu2008a}, on both types of graphs. The results agree on the
higher quality of ranking based on the browse graph, because it is a more
reliable source; they also tried out a combination of the two graphs with very
interesting outcomes.
The user browsing graph and related PageRank-like algorithms have been exploited
to rank different types of items including
images~\cite{trevisiol12image,chiarandini12discovering},
photostreams~\cite{chiarandiniICWSM13}, 
and predicting users
demographic~\cite{hu07demographic} or optimizing web crawling~\cite{liu11user}.
Trevisiol \etal~\cite{trevisiol12image} made a comparison between different
ranking techniques applied to the Flickr \bgraph.
Chiarandini \etal~\cite{chiarandini12discovering} found strong correlations between the type of user's navigation and the type of external Referrer URL.
Hu \etal~\cite{hu07demographic} have shown that demographic information of the users (\eg, age and gender) can be identified from their browsing traces with good accuracy.
The \bgraph\ has been used also for recommending sequences of photos that users often like to navigate in sequence, following a collaborative filtering approach~\cite{chiarandiniICWSM13}.
In order to implement an efficient news recommender the user's taste have to be considered as they might change over time. Indeed, studying the users browsing patterns, Liu \etal~\cite{liu10personalized} showed that more recent clicks have a considerably higher value to predict future actions than the historical browsing record. 
Finally, Trevisiol \etal~\cite{TrevisiolRecSys14} exploited the \bgraph\ in order to build some user models in the news domain, and recommend the next article the user is going to visit. They introduced the concept of \rgraph, which is a \bgraph\ built with sessions that are generated by the same referrer domain. 
Even if the purposes of our work are very different, we construct the \rgraphs\ in the same way in order to be in-line with their investigation. 

To the best of our knowledge there is no work in the state of the art that tackles the {\it Local Ranking Problem} on a browsing graphs with the prediction task that we perform and describe in this paper.

\section{Dataset} \label{sec:dataset} 
For the purpose of this study, we took a sample of Yahoo News network's\footnote{We considered a number of different subdomains like {\it Yahoo news}, {\it finance}, {\it sports}, {\it movies}, {\it travel},
{\it celebrity}, \etc} user-anonymized log data collected in $2013$.
The dataset used in this work has been extracted from the data built in~\cite{TrevisiolRecSys14}, that was used with the purpose to study the news consumption with respect to the Referrer URL.
In this section we summarize how we built the dataset and the graphs, but the reader may refer to the aforementioned paper for further details.
The data is comprised by a large number of pageviews, which are represented as
plain text files that contain a line for each HTTP request satisfied by the Web
server. For each pageview in the dataset, we gathered the following fields:
\begin{equation*}
\left\langle BCookie, Timestamp, ReferrerURL, URL, User Agent\right\rangle
\end{equation*}
The \textit{BCookie} is an anonymized identifier computed from the browser
cookie. This information allowed us to re-construct the navigation session of
the different users. \textit{URL} and \textit{ReferrerURL} represent, respectively, the current page the user is visiting and the page the user visited before arriving at the destination page.
Note that the \textit{Referrer URL} could belong to any domain, \eg, it may be
external to the Yahoo News network. 
The \textit{User-Agent} identifies the user's browser, an information that we used to filter out Web crawlers,
and \textit{Timestamp} indicates when the page was visited. 
All the data were anonymized and aggregated prior to building the browsing graphs.
We removed traffic derived from Web crawlers by preserving only the entries whose User-Agent field contains a well-known browser identifier.
After applying the filtering steps described above, our sample contains approximately $3.8$ million unique pageviews and $1.88$ billion user transitions.

\subsection{Session Identification and Characteristics} \label{sec:sessions}
The \bgraph\ is a graph whose nodes are web pages, and whose edges are the
browsing transitions made by the users. To build it we extract the
transitions of users from page to page, and in order to preserve the user
behavior (that could vary over time), we group pageviews into \emph{sessions}.
We split the activity of a single user, taking the \textit{BCookie} as an
identifier, into different sessions when either of these two conditions holds:
\begin{itemize}
  \item {\bf Timeout:} the inactivity between two pageviews is longer than $25$ minutes.
  \item {\bf External URL:} if a user leaves the news platform and returns from an external domain, the current session ends even if previous visits are within the 25 minute threshold.
\end{itemize}
Moreover, each news article of the dataset is annotated with a high-level {\it category} manually assigned by the editors.

\subsection{Subgraphs Based on Session Referrer URL} \label{sec:subgraphs}
%
\begin{table}[!ht]
\centering
  \setlength{\tabcolsep}{.7em}
  \begin{tabular}{@{}lrrrrrrr@{}} \toprule
    \textbf{Subgraphs} & \textbf{Nodes} & \textbf{Edges} & \textbf{Density} & \textbf{\%GCC} \\
    \midrule
    Google       & $142,646$    & $779,185$ & $3.8\cdot10^{-5}$ & $0.93$ \\
    Yahoo       & $101,116$   & $404,378$ & $3.9\cdot10^{-5}$ & $0.95$ \\
    Bing         & $61,531$   & $255,464$ & $6.7\cdot10^{-5}$ & $0.91$ \\
    Homepage    & $60,287$    & $335,836$ & $9.2\cdot10^{-5}$ & $0.99$ \\
    Facebook     & $21,060$     & $70,266$  & $1.5\cdot10^{-4}$ & $0.95$ \\
    Twitter         & $4,206$     & $7,080$   & $4.0\cdot10^{-4}$ & $0.87$ \\
    Reddit       & $2,445$      & $4,868$   & $8.1\cdot10^{-4}$ & $0.95$ \\
    \bottomrule
  \end{tabular}
  \caption{Size, density and Giant Connected Component of the extracted subgraphs. Note that there is not a strict relation between the size of the subgraph and the amount of browsing traffic generated in it.}
  \label{tab:infosubgraphs}
\end{table}
We aim to compare the PageRank scores of the nodes between the full \bgraph, computed with all the Yahoo News logs, and a subgraph that represents the local graph. This is a way to simulate a real-world scenario in which a service provider knowns only the users navigation logs inside its network (subgraph) while the external navigations are unknown (full \bgraph). 
Since it is not possible to use the full Web browsing log, we perform a simulation using different subgraphs extracted from the same \bgraph\ that represent the local graphs of different providers.
One possible approach would be to simulate each service provider as a different Yahoo News subdomain (\eg, news, sports, finance). However, news articles are often shared on different Yahoo subdomains and, as a consequence, the users jump among different subdomains in each single session. To avoid such an overlap on the subgraphs, we define a different simulation approach.
We extract from the \bgraph\ of the Yahoo News dataset various subgraphs built with sessions of users generated by the same Referrer URL. It has been shown~\cite{TrevisiolRecSys14} that \bgraphs\ constructed in this way contain very different users sessions in terms of content consumed (nodes visited).
In particular we consider users accessing the news portal directly from the homepage, which is the main entry point for regular news consumption, and in addition from a number of domains that fall outside the Yahoo News network: {\it search engines} (Google, Yahoo, Bing) and {\it social networks} (Facebook, Twitter, Reddit).
For each source domain we extract a subgraph from the overall \bgraph, by considering only the browsing sessions whose initial {\textit Referrer URL} matches that domain. For example, if a user clicks on a link referring to our network that has been posted on Twitter, her {\textit Referrer URL} will be the Twitter page where she found the link. Next, we consider all the following pageviews belonging to the same session of the user, as being a part of the {\it twitter-subgraph}, given that all of them have been reached through Twitter.
We applied the same procedure for all the sources defined before, and finally, we obtained a weighted graph for each different external URL. The \textit{Weight} accounts for the number of times a user has navigated from the source page to the destination page. On Table~\ref{tab:infosubgraphs} a summary with the size of the graphs (in terms of number of nodes and edges) and with their structure is shown. It is interesting to see that all the graphs, even presenting very different size, are very well connected (\%GCC between $0.87$ and $0.99$).

\section{Referrer Graphs Analysis} \label{sec:analysis} 
%
\begin{table*}[!ht]\centering
  \begin{tabular}{@{}lcccccccc} \toprule
  & \textbf{Full} & \textbf{Facebook} & \textbf{Google} & \textbf{Bing} & \textbf{Yahoo}  & \textbf{Reddit} & \textbf{Homepage} & \textbf{Twitter}  \\
  \midrule
  Full& \cellcolor[gray]{0.40}1.0000  & \cellcolor[gray]{0.89}0.1791  & \cellcolor[gray]{0.76}0.3931  & \cellcolor[gray]{0.80}0.3278  & \cellcolor[gray]{0.79}0.3548  & \cellcolor[gray]{0.96}0.0656  & \cellcolor[gray]{0.83}0.2797  & \cellcolor[gray]{0.95}0.0764  \\
  Facebook& \cellcolor[gray]{0.89}0.1791  & \cellcolor[gray]{0.40}1.0000  & \cellcolor[gray]{0.81}0.3146  & \cellcolor[gray]{0.75}0.4111  & \cellcolor[gray]{0.79}0.3430  & \cellcolor[gray]{0.84}0.2616  & \cellcolor[gray]{0.76}0.4070  & \cellcolor[gray]{0.82}0.3026  \\
  Google& \cellcolor[gray]{0.76}0.3931  & \cellcolor[gray]{0.81}0.3146  & \cellcolor[gray]{0.40}1.0000  & \cellcolor[gray]{0.65}0.5815  & \cellcolor[gray]{0.65}0.5860  & \cellcolor[gray]{0.93}0.1088  & \cellcolor[gray]{0.75}0.4217  & \cellcolor[gray]{0.92}0.1297  \\
  Bing& \cellcolor[gray]{0.80}0.3278  & \cellcolor[gray]{0.75}0.4111  & \cellcolor[gray]{0.65}0.5815  & \cellcolor[gray]{0.40}1.0000  & \cellcolor[gray]{0.60}0.6624  & \cellcolor[gray]{0.91}0.1469  & \cellcolor[gray]{0.69}0.5238  & \cellcolor[gray]{0.90}0.1688  \\
  Yahoo& \cellcolor[gray]{0.79}0.3548 & \cellcolor[gray]{0.79}0.3430  & \cellcolor[gray]{0.65}0.5860  & \cellcolor[gray]{0.60}0.6624  & \cellcolor[gray]{0.40}1.0000  & \cellcolor[gray]{0.93}0.1245  & \cellcolor[gray]{0.72}0.4632  & \cellcolor[gray]{0.92}0.1386  \\
  Reddit& \cellcolor[gray]{0.96}0.0656  & \cellcolor[gray]{0.84}0.2616  & \cellcolor[gray]{0.93}0.1088  & \cellcolor[gray]{0.91}0.1469  & \cellcolor[gray]{0.93}0.1245  & \cellcolor[gray]{0.40}1.0000  & \cellcolor[gray]{0.91}0.1534  & \cellcolor[gray]{0.86}0.2309  \\
  Homepage& \cellcolor[gray]{0.83}0.2797  & \cellcolor[gray]{0.76}0.4070  & \cellcolor[gray]{0.75}0.4217  & \cellcolor[gray]{0.69}0.5238  & \cellcolor[gray]{0.72}0.4632  & \cellcolor[gray]{0.91}0.1534  & \cellcolor[gray]{0.40}1.0000  & \cellcolor[gray]{0.91}0.1523  \\
  Twitter& \cellcolor[gray]{0.95}0.0764 & \cellcolor[gray]{0.82}0.3026  & \cellcolor[gray]{0.92}0.1297  & \cellcolor[gray]{0.90}0.1688  & \cellcolor[gray]{0.92}0.1386  & \cellcolor[gray]{0.86}0.2309  & \cellcolor[gray]{0.91}0.1523  & \cellcolor[gray]{0.40}1.0000  \\
  \bottomrule
  \end{tabular}
  \caption{Kendall's $\tau$ correlations between PageRank values ($\alpha=0.85$) between the common nodes of the subgraphs.}
  \label{tab:kendallsubgraphs}
\end{table*}
In this section we describe some analysis on these \rgraphs, proving that they are consistently different not only in term of nodes and content
but also in term of navigation patterns of the users. We also propose an experiment to understand how much the graphs are distinguishable.
\vspace{.4cm}

\subsection{Subgraphs comparison} \label{sec:analysis:pagerank} 
We consider the seven subgraphs extracted from the main news portal graph with the procedure discussed in $\S$\ref{sec:dataset}. 
Browsing patterns generated by different types of audience, can lead to different pieces of news pages to emerge as the most central ones in the \browsegraph. 
To check that we ran the PageRank algorithm on each of the (weighted) subgraphs, and for every pair of subgraphs we compared the scores obtained on their common nodes using Kendall's $\tau$ distance. 
The intersection between the node sets of the networks is always large enough to allow us to compute the $\tau$ on the intersection only ($>1000$ nodes in the case with less overlap). Kendall's $\tau$ will provide a clear
measure of how much the importance of the same set of nodes varies among different subgraphs. 
When the ranking between two subgraphs differs greatly (\ie, low Kendall's $\tau$), it is an indication that they either show different content (\ie, webpages) or that the collective browsing behaviour in the two graphs privileged different sets of pages.

Table~\ref{tab:kendallsubgraphs} reports on the cross-distance among the
subgraphs and also with respect to the full graph using Kendall's $\tau$.
Interestingly, most of the similarity values tend to be very low (<0.3),
confirming the hypothesis that the user's interests are tightly related to the
domain where they come from. Some of these similarities, however, are
considerably higher, remarkably the ones between the three subgraphs that are
originated from search engines traffic, \ie, Bing, Google and Yahoo, which yield
the most similar rankings of pages (>0.5).
However, for the purpose of this work we expect to find a difference among the subgraphs in order to use them as local \bgraph\ and study the LRP with the full graph (\ie, \bgraph\ made with the entire news log).

\subsection{Random Surfer} \label{sec:randomsurfer}
In $\S$\ref{sec:analysis:pagerank} we showed how users coming from different sources (\ie, referrer domains) behave differently in terms of content discovery and, as a consequence, the importance of the news articles vary significantly among the different \bgraphs.
It has been shown how the referrer domain might be extremely useful to characterize user sessions~\cite{chiarandini12discovering}, to estimate user engagement~\cite{lehmann2013measuring} or to perform cold-start recommendation~\cite{TrevisiolRecSys14}.
However, the user's referrer URL is not always visible and, in many cases, it is hidden or masked by services or clients. For instance, any Twitter or mail client (\ie, third-party application) shows an empty referrer URL in the web logs. A similar situation happens with the widespread URL-shortening services (\eg, \texttt{Bitly.com}), which mask the original Web page the user is coming from. Nonetheless, in all these cases a provider could make use of her knowledge of the user's trail, to identify automatically the source where the user started her navigation in the local graph. As we have shown, the referrer URL might be useful to characterize the interest of the users, especially in the case where the users are unknown (\ie, the user profile is not available). Thus, being able to identify the referrer URL when it is not available, is an advantage for the content provider. In this section we want to understand if it is feasible to detect the referrer URL of a user while he browses and how many browsing steps are required to be able to do so accurately. 
\begin{algorithm}[!th]
  logPr $\leftarrow$ initialize ~vector ~with ~size ~$G_k.length()$\;
  n $\leftarrow$ total number of nodes\;
  $x_j \leftarrow$ choose (random) starting node $\in G_k$\;
  \BlankLine
  \emph{/* For each step, compute a random walk in $G_k$, and compare the probability to be in all the other $G$ */}\\
  \For {$s \leftarrow 1 \TO steps$} {
    \BlankLine
    \emph{/* Pick the next node of $G_k$ with random walk */}\\
    $x_k$ = next\_node( $G_k$, $x_j$ )\;
    \BlankLine
    \For {$i\leftarrow 0$ \KwTo $G.length()$} {
      $\left\langle k_{out}\right\rangle \leftarrow$ \texttt{get\_outdegree}($n_p$)\;
      \eIf {$\left\langle k_{out}\right\rangle == 0$} {
        logPr[ i ] $\leftarrow$ logPr[ i ] + $\log( 1 / n )$\;
      }{
        $p_i(x) = (1-\alpha) / n$\;
                  $Pd_{x_j} \leftarrow$ \texttt{get\_prob\_distribution}$(G_i,~x_j)$\;
                  $S_{x_j} \leftarrow$ \texttt{get\_successors}$(G_i,~x_j)$\;
        \If {$x_k \in S_{x_j}$} {
          $p_i(x) \leftarrow p_i(x) + \alpha * Pd_{x_j}(x_k)$\;
        }
        logPr[ i ] $\leftarrow$ logPr[ i ] + $\log( p_i(x) )$\;
      }
    }
  }
  \Return{}~logPr
\caption{\texttt{RandomSurfer}(k,~$\alpha$,~steps,~G)}\label{algo:randomsurfer}
\end{algorithm}

Moreover, if the user sessions are easily distinguishable it means that the subgraphs are different enough to be considered, in our experiment, as \emph{local} \bgraphs\ of different service providers.

Therefore, we consider the following scenario: a content provider is observing a user
surfing the pages of its web service, but it is unaware of the user's referrer URL. 
In terms of our experimental dataset, this scenario maps into the problem of observing a browsing trace left by a random surfer on one of the referrer-based subgraphs and having to identify which graph it is. Intuitively, the larger the number of page visits (or \textit{steps}) the surfer will make the more distinctive its trace will be, and the easier the identification of the graph. Algorithm~\ref{algo:randomsurfer} shows the pseudocode that describes the process to compute the random surfer experiment.

Formally, observing the sequence of the surfer's visited nodes $\mathbf{x}=(x_1, x_2, \dots,
x_s)$ and computing the probability $p_i(\mathbf{x})$ that the surfer has gone
through them given that it is surfing $G_i$, we need to deduce what is
$G_i$ (\eg, by maximum log-likelihood). With this goal in mind, we sort the indices of
the subgraphs $i_1,i_2,\dots$ so that $p_{i_1}(\mathbf{x})\geq
p_{i_2}(\mathbf{x}) \geq \dots$ and stop as soon as the gap between $\log
p_{i_1}(\mathbf{x})$ and $\log p_{i_2}(\mathbf{x})$ is large enough (\eg, $\log
p_{i_1}(\mathbf{x})-\log p_{i_2}(\mathbf{x}) \geq \log 2$), 
with a maximum of $20$ steps that we consider as a representation of a long user session. \\
\begin{figure}[!t]
\centering
  \includegraphics[clip=true, width=\columnwidth]{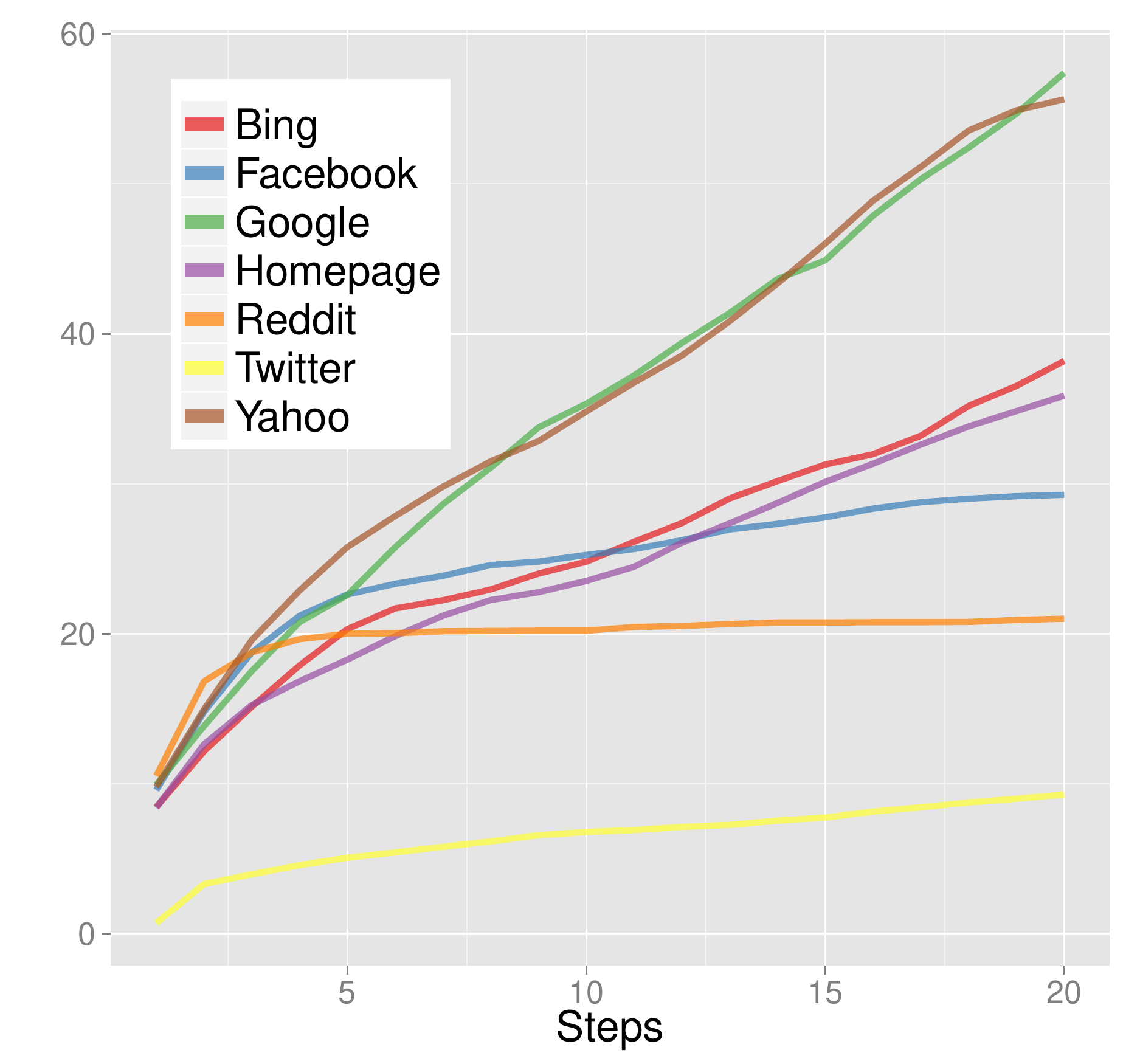} 
  \caption{Random Surfer Experiment. Y-axis: log-ratio of the probabilities between the correct graph and the graph with the largest log-probability (as explained in the text).
  X-axis: number of browsing steps performed by the surfer.}
  \label{fig:random-surfer}
\end{figure}
In this set of experiments, we considered the seven URL-referral subgraphs $G_1$, \dots, $G_7$, one at a time.
For each subgraph $G_i$, we simulated a random
surfer moving around in $G_i$ (\ie, calling the function \texttt{RandomSurfer}(i, $\alpha$, steps, G)), computing at each step (\ie, page visited) the probability of the surfer to navigate in each subgraph $G_1$, \dots $G_7$: 
we expect that the probability corresponding to $G_i$ will increase at each step, and will 
eventually dominate all the others.

To estimate the number of steps required to identify correctly the graph that
the surfer is browsing, we measure the difference between log-probabilities for
the correct graph $G_i$ and for the graph with the largest log-probability among the
other ones.
As with PageRank we introduced a certain damping factor
($\alpha=0.85$); this is necessary to avoid being stuck in terminal components
of the graph. Recall that $\alpha$ is the balancing parameter that determines
the probability of following in the random walk, instead of teleporting. The
results are shown in Figure~\ref{fig:random-surfer}, averaged over $100$ executions. 
The values on the y-axis represent the difference between the log-probabilities 
(\ie, the logarithm of their ratio): in general, we can observe that
the very first steps are enough to understand correctly (and with a huge margin) in which
graph the surfer is moving. The inset of Figure~\ref{fig:random-surfer} displays
the first $20$ steps and the relative probability to identify the correct graph.
Almost all the referrer domains are recognizable at the first step. This translates into a strong advantage for the service provider as it can identify from where the users are coming from, even if they use clients or services that masquerade it. With this information the service provider can personalize the content of the web pages for any users with respect to the referrer.

Interestingly, the plot reveals that some surfers are
easier to single out than others; we read this as yet another confirmation that
the subgraphs have a distinguished structural difference, or (if you prefer) that
users have a markedly different behavior depending on where they come from.
This experiment does not only showed that is possible to detect from which referrer domain the surfer is coming from, but that the graphs are quite different and that they can be used for our study.

\section{PageRank on the BrowseGraph} \label{sec:lrp-analysis} 
Next, we study the convergence of the PageRank ranking between the \emph{local} \bgraphs\ (\rgraphs) and the full \bgraph. We want to understand how different are the ranking computed using less or more knowledge about the full graph. We present an experiment, called ``Growing Rings'', which compute the distance between the rankings expanding at each step the known nodes (and edges) with the neighbors of the subgraphs.

\subsection{``Growing Rings'' Experiment} \label{sec:growingrings}
We first focus on the study of the {\it Local Ranking Problem} on browsing
graphs. An important question related to this problem is how much the PageRank
node values vary, when new nodes and edges are added to the local graph. A
natural way to determine this is to expand incrementally the graph by adding new
nodes and edges in a Breadth-First Search (BFS) fashion, and comparing the
PageRank computed on the expanded graph with the one on the global graph.

More formally, given a graph $H$ which is a subgraph of the full graph $G$, we
simulate a growth sequence $H_0$, $H_1 \dots H_n$ in the following way:
\begin{itemize}
  \item $H_0 \longleftarrow H$;
  \item $V_{H_{k+1}} \longleftarrow \{\Gamma_{out}(V_{H_k}) \cup V_{H_k}\}$, with $V_x$ being the set of vertices of a graph, and $\Gamma$ being the vertex neighborhood function;
  \item $E_{H_{k+1}} \longleftarrow \{(v_1, v_2) | v_1 \in V_{H_{k+1}} \wedge v_2 \in V_{H_{k+1}} \}$, with $E_x$ being the set of edges of a graph.
\end{itemize}
We refer to the various steps of this
expansion as \textit{``rings''}, where the ring $H_0$ is the initial subgraph
and subsequent rings are obtained by adding all the outgoing arcs that depart
from the nodes in the current ring and end in nodes that are not in the ring. Observe that, depending on how it is built, $H_0$ may not be
an induced subgraph of $G$, but $H_1, \cdots, H_n$ are always induced subgraphs, by
definition of the expansion algorithm.

Using the Kendall's $\tau$ function, we measure the difference between the local
PageRank computed for each ring $H_i$ and the global PageRank computed on $G$.
The main objective is to understand how much the ranking gets close the global
one at each consecutive step, and whether the ranking values are able to
converge even if we just consider a piece of the information contained in the
whole graph.

To check the dependency of results from the initial graph selected, we consider
three different sets of initial subgraphs, which we will study separately. We
describe them next.
\begin{itemize}
\item \textbf{Referrer-based (RB)}. The seven browsing subgraphs built by referrer URL: Facebook, Twitter, Reddit, Homepage, Yahoo, Google and Bing;
\item \textbf{Same size referrer-based (SRB)}. To measure how much the different sizes of the graphs impact on the observed behavior, we fix a number of nodes and extract a portion of each subgraph in order to obtain exactly the same size for all networks. The selection is performed with several attempts of BFS expansion, starting from a random node in each graph, until the resulting graphs have very similar size ($\pm 9.4\%$): other ways of selecting subgraphs would end up with disconnected samples, which of course would void the purpose of this experiment. With this procedure instead, we are able to compare the graphs on equal grounds and at the same time control for the effect of size (about $3K$ nodes and $20K$ edges).
\item \textbf{Random (R)}. To check whether the observed behavior has to do with the user behavior underlying the graph under examination (\textit{e.g.}, the particular structure of the graph determined by the sessions of users coming from Twitter), we take a set of seven \textit{random} graphs each of them reflecting the size of each of the referrer-based subgraphs. 
Thus, we can explore the behavior of browsing graphs, which preserve the size of the graphs originated by specific types of users, but that are ``artificial'' in the sense that destroy any connection with the behavior connected to a particular user class. To make sure that the size is the same, we start from a BFS exploration and we prune the last level to match exactly the size we need.
\end{itemize}

\begin{figure*}
   \includegraphics[width=\textwidth]{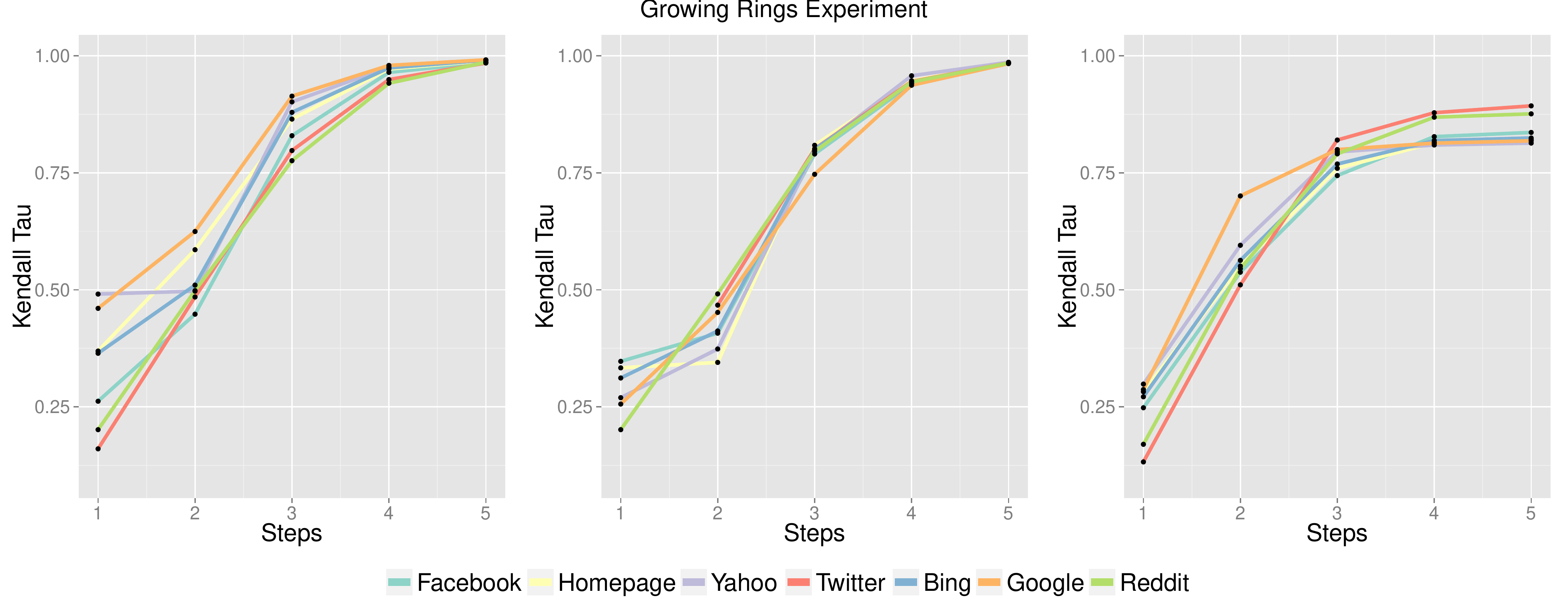}
   \caption{Growing Rings experiment on: (left) original subgraphs built based on the referrer URL, (center) seven subsubgraphs with very similar size, (right) seven subgraphs random selected from the full graph, where each of them has the same size of one of the original.}
   \label{fig:growinrings}
\end{figure*}

The results related to the \textbf{RB} case are shown in
Figure~\ref{fig:growinrings} (left). The convergence happens relatively quickly,
as the value $\tau$ approaches $1$ in the first $3$ iterations. The curves
related to different subgraphs are shifted with respect to each other,
apparently mainly due to their different size, the biggest networks starting
from higher $\tau$ values and converging faster than the smaller ones.
To determine the dependency on the graph size, we repeat the same experiment for the \textbf{SRB} case. The results for this case are shown in Figure~\ref{fig:growinrings} (center). Even if the curves resulted to be more flattened (confirming that the initial size has indeed a role in the convergence), we still observe noticeable differences between the curves for the first two expansion levels. This means that different subgraphs are substantially different from one another in terms of their structure: even after forcing them to have the same size, the convergence rates observed on the different graphs varies. At the first iteration, for instance, all the subgraphs in \textbf{SRB}  have Kendall's $\tau$ between $0.3$ and $0.5$, whereas the ones in \textbf{RB} are between $0.4$ and $0.6$.
Moreover in \textbf{SRB} the biggest networks starting from higher $\tau$ values are not converging faster. This intuition is confirmed by repeating the experiment on graphs selected with the \textbf{R} strategy. Results, displayed in Figure~\ref{fig:growinrings} (right), show that convergence in this case is much slower and the difference between the curves is less prominent. 

Summarizing, with the previous experiment we show that the Growing Rings on random subgraphs behave differently, especially when considering the number of iterations required in order to converge.

\subsection{Growing Rings with Selection of Nodes}
Besides the selection of the initial graph, the rank convergence depends also on
the way the growing rings are built at each iteration. How does the expansion
influence convergence if only few more representative nodes are selected? To
what extent a higher \textit{volume} of selected nodes helps a quicker
convergence or adds more {\it noise}? At a first glance, one may argue that
using all the nodes is equivalent to injecting all the available information, so the
convergence to the values of PageRank computed on the full graph $G$ should be
faster. On the other hand, instead, one may observe that we are introducing a
huge number of nodes in each iteration (as the growth is at each step larger),
adding also the ones that are less important and this can induce an incorrect
PageRank for some time, until all the graph becomes known.
In order to shed light on this aspect, we introduce a variant in the
growing-rings expansion algorithm and we select only the nodes with highest
PageRank. 

More formally, considering $H_k$ as the subgraph at iteration $k$ and $V_{H_k}$
its set of nodes, we select all the external nodes in $Y = \{V_G \backslash V_{H_k}\}$,
which are connected through outgoing arcs from the nodes in $V_{H_k}$.
We then compute the PageRank values on the subgraph $H_k$ extended with the
nodes $Y$ and obtain a ranked list of nodes. Among all the nodes in $Y$ we
select the top $n\%$ with largest PageRank value, and only those ones will be
added to $H_k$ in order to build $H_{k+1}$ and advance to the next iteration.

\begin{figure}[!th]
\centering 
  \includegraphics[width=\columnwidth]{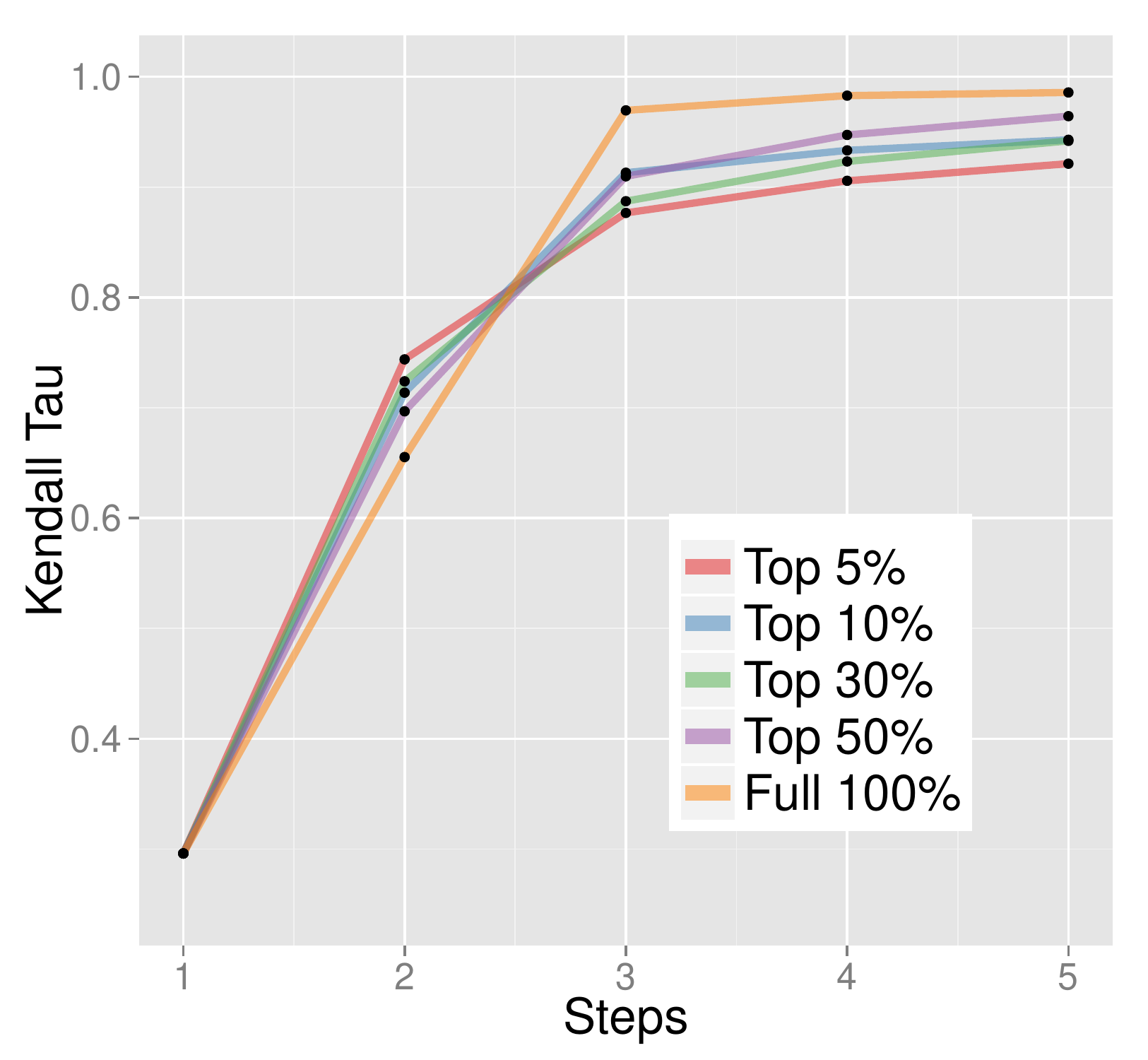}
\caption{Growing Rings using only the nodes with highest PageRank. The plot shows the average values of the Kendall-$\tau$ at each step computed for all the subgraph.}
\label{fig:gb-partial}
\end{figure}

We conducted experiments with this partial expansion at different percentages:
$5\%$, $10\%$, $30\%$, $50\%$, and $100\%$, and then we computed the
average Kendall's $\tau$ value for each one of the percentages. The results are 
shown in Figure~\ref{fig:gb-partial}. Remarkably,
the figure highlights how expanding the graph by adding fewer nodes, although
the most representative ones, leads to PageRank values that are closer to the
\emph{global} ones in the first iterations. Since we are expanding the local
graph with a small (highly-central) number of nodes, we could argue that they
initially help to boost the local PageRank scores. However, given that we keep
on expanding using a few nodes at each iteration, the nodes that have not been
added before exclude a large number of nodes among which there might also be
highly central ones. This might explain why in the first iteration(s) the
convergence rate is high, but on the limit the final convergence values result in a
low Kendall's $\tau$.
Contrarily, in the long run, expansions that include the highest number of nodes
present convergence values closer to $1$. This is somehow expected, given that
at each iteration any subgraph $H$ closer in size to the full graph $G$ will
include almost every node and arc. 

Nonetheless, the main significant outcome of this experiment is that it is
possible to obtain a yet satisfactory PageRank convergence, with few but very
representative nodes. For situations in which including additional
pieces of information (in terms of node/arc insertions) implies a non-negligible
cost, requesting just a little amount of well-selected information allows to
obtain good approximations while minimizing the costs.

\section{PageRank Prediction} \label{sec:prediction} 
In the previous section we have shown how the approximation to the global PageRank varies
with the expansion of the initial subgraph. 
The ranking of the nodes converges quite fast on all the subgraphs: they differ
in terms of their content, although they are similar in terms of structure in that all of them are built based on users' navigational patterns.
Building upon the findings about how local and global PageRank computed on the \bgraphs\ relate to each other, we designed an experiment to assess how well a learned model could perform in predicting this relationship.

We address the problem of predicting the Kendall's $\tau$ between the local and the global PageRank, only considering information available on the local graph such as topological features. This is an extremely common situation given that, in general, the information pertaining the local graph is the only one that is readily available and usually of a limited size.
Computing this distance accurately has a high value for service providers, since it translates directly into an estimation of the reliability of the PageRank scores computed on their local subgraphs.
As a direct consequence one can apply, with different levels of confidence, methods for optimizing web sites~\cite{WeischedelWOW2006}, studying user engagement~\cite{lehmann2013measuring}, characterizing user's session~\cite{chiarandini12discovering} or content recommendation~\cite{TrevisiolRecSys14}.

\subsection{Prediction of Kendall $\tau$ Distance} \label{sec:deltaprediction}
We have seen that the deviation of the local PageRank 
with respect to the global one can be relevant, depending on
factors such as the size of the local graph and the different behavior of the
users who browse it (see $\S$\ref{sec:growingrings} and particularly
Figure~\ref{fig:growinrings}). Recall that we compute the distance comparing the rankings with Kendall's $\tau$, since we are interested in obtaining a ranking as close as possible to the one computed on the entire graph.
Although we have previously shown how to expand the view on the local graphs with nodes residing at the border, this practice might not always be possible in a real-world scenario, since service providers often can have access only to the browsing data \emph{within} their domain.

Previous work on local ranking on graphs raised several questions related to
this scenario, highlighting practical applications of the local rank estimation
non only for web pages but also in social networks~\cite{Bressan2011}.
Critically, so far it is not clear whether there are some topological properties
of the local graph that make the local ranking problem easier or harder, and if
these properties can be exploited by local algorithms to improve the quality of
the local ranking. We explore this research direction by studying a fundamental
aspect that is at the base of the open questions in this area, namely the
possibility of estimating the deviation of the local PageRank from the global
one, using the structural information of the local network. The intuition is that,
some structural properties of the graph could be good proxies for the $\tau$
value difference, computed between local and global ranks.
Being able to estimate the Kendall's $\tau$ distance between the subgraph available to the service provider and the global graph, implies the ability to estimate the reliability of the current ranking using only information of the local subgraph.

\begin{figure}[!t]
\centering
  \includegraphics[clip=true, width=\columnwidth]{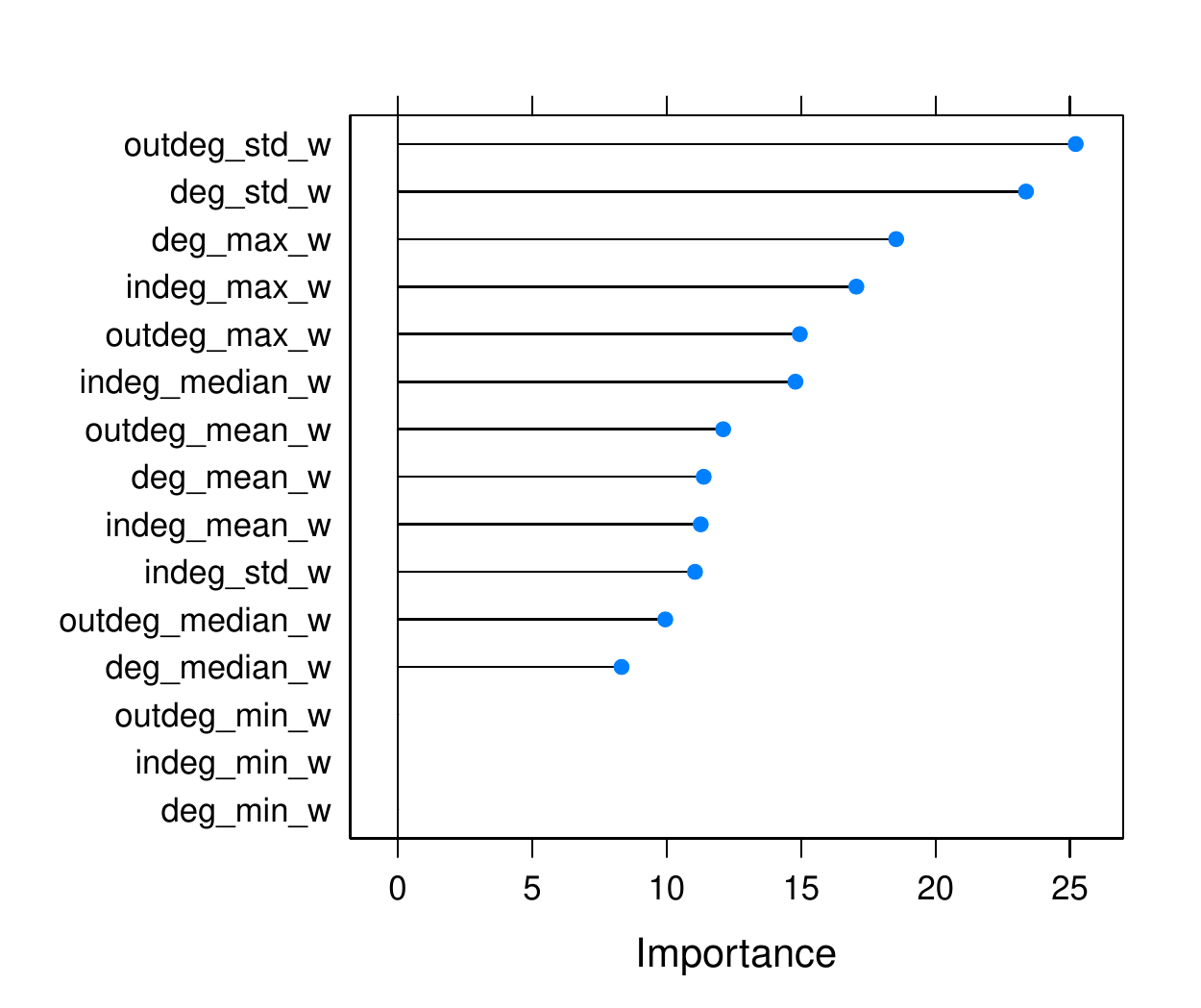} 
  \caption{The $15$ features of \emph{weighted degree}, the most predictive class, sorted by importance. Note that some of them do not have any contribution to the Kendall-$\tau$ prediction, therefore just few features are necessary in order to estimate the distance.}
  \label{fig:feat-importance}
\end{figure}

To verify this hypothesis we resort to regression analysis. Starting from the
seven subgraphs in the dataset, we build a training set using the jackknife
approach, by removing nodes in bulks ($1\%$, $5\%$, $10\%$, $20\%$) and
computing the $\tau$ value between the full subgraph and their reduced versions.
Then, for each instance in the training set we compute $62$ structural graph
metrics~\cite{wasserman94social,Barrat2008} belonging to the following
categories:

\begin{itemize}
\item \textbf{Size and connectivity (S)}. Statistics on the size and basic wiring
properties, such as number of nodes and edges, graph density, reciprocity,
number of connected components, relative size of the biggest component.
\item \textbf{Assortativity (A)}. The tendency of node with a certain degree, to be
linked with nodes with similar degree. We computed different combinations that
take into account the in/out/full degree of the target node vs. the in/out/full
degree of the nodes that are connected with it.
\item \textbf{Degree (D)}. Statistics (average, median, standard deviation, \etc) on
the degree distribution of nodes.
\item \textbf{Weighted degree (W)}. Same as \textbf{degree}, but considering the
weight on edges, which usually referred as node strength. As the edges are the
transitions made by the users during the navigation, the weight stand for the
number of times the users have navigated the transition.
\item \textbf{Local Pagerank (P)}. Statistics on the distribution of the PageRank
values computed on the local graph.
\item \textbf{Closeness centralization (C)}. Statistics on the distances (number of
hops), which separate a node to the others in the graph, in the spirit of the
closeness centralization~\cite{wasserman94social}.
\end{itemize}
We employed different regression algorithms, although we report the performance
using random forests~\cite{Breiman:2001:RF:570181.570182}, which performed
better in this scenario than other approaches like support vector
regression~\cite{Smola03atutorial}. We computed the mean square error (MSE)
across all examples in all sampled subgraphs. 
The random forest regression has been computed over a five-fold cross validation averaged over $10$ iterations.
The mean square residuals that we obtained is very low, around $2.4\cdot10^{-6}$.
Results, computed for the full
set of features and for each category separately, are given in
Table~\ref{tab:mse}. The most predictive feature category is the \emph{weighted degree}, which yields a performance that is better (or comparable) than the model
using all the features. The \emph{assortativity} features instead, seem to be the ones that have the less predictive power on their own.
This might be due to the fact the model with 62 features is too complex for the amount of training data available. On the other hand, the \emph{weighted degree} that is the best performing class of features, contains the statistics of the degree distribution on the weighted edges. 
In Figure~\ref{fig:feat-importance} the features included in \emph{weighted degree} are ranked by their discriminative power in predicting the Kendall $\tau$ distance using the permutation test proposed by Strobl \etal~\cite{Strobl2008rf}. These features, which are based on the distribution of the out- and in-degree of the nodes, are straightforward to compute from the local graph---a very affordable task for service providers.

\begin{table}[!tbp]\centering
  \setlength{\tabcolsep}{1.5em}
  \begin{tabular}{@{}lcl@{}} \toprule
    \textbf{Feature Class} & \textbf{No. Features} & \textbf{MSE}   \\
        \midrule 
        weighted degree       & $15$  & $2.2\cdot 10^{-6}$\\ 
          degree            & $15$  & $2.9\cdot 10^{-6}$\\ 
          local PageRank      & $10$  & $3.3\cdot 10^{-6}$ \\ 
        size and connectivity   & $9$   & $3.4 \cdot 10^{-6}$  \\ 
          closeness           & $5$   & $4.1\cdot 10^{-6}$\\ 
          assortativity         & $8$   & $9.3\cdot 10^{-6}$ \\ 
          \midrule
        ALL features        & $62$  & $2.4\cdot10^{-6}$\\ 
    \bottomrule
  \end{tabular}
  \caption{MSE of cross validation. Average differences are statistically significant with respect to \emph{weighted degree} and \emph{ALL features} (t-test, p<0.01).}
  \label{tab:mse}
\end{table}

We then use the learned model to predict the $\tau$ values of the seven
subgraphs. When we applied the predictive models learned in the subsamples to
regressing the full graphs, the MSE, is less than 0.026 on average, which, even
if relatively low, it is higher than the cross-validated performance in the
sub-samples. However, the model was able to rank the seven different subgraphs
by their Kendall's $\tau$ almost perfectly. When using all the features the
Spearman's correlation coefficient between the true order and the predicted one
is 0.85 (high correlation), and when we used the most predictive features
(weighted degree) the correlation was as high as 0.80 (moderate high
correlation). Overall, the final rankings are just one swap away (Kendall's
$\tau$ is over 0.70 in this case).
This kind of information can be very helpful when comparing different local
sub-domains to determine which one has pages that better estimate the
global PageRank.

\section{Conclusion} \label{sec:conclusions}
In this paper we tackled the {\it Local Ranking Problem}, \ie, how to estimate
the PageRank values of nodes when a portion of the graph is not available, which
arises commonly in real use cases of PageRank. We investigated this problem for
a novel environment, namely estimating PageRank on a large user-generated browsing
graph from Yahoo News.
The peculiar characteristic of this graph is that it is
built from user's navigation patterns, where nodes represent
web pages and edges are the transitions made by the users themselves. Moreover, the
information about the domain of origin of the users (namely the referrer URL of
their sessions), is also available.

We built a set of \textit{ReferrerGraphs} including the browsing subgraphs based on different referrer URLs, and then we studied their difference in terms of user navigation patterns. 
We found that all of the browsing patterns initiated from different domains exhibit remarkable differences in terms of which pages users visited next. 
The referrer URL (or domain) has been found to be extremely useful for characterizing the user behavior~\cite{chiarandini12discovering} or for recommendation of content~\cite{TrevisiolRecSys14}.
With this observation in mind and motivated by the cases where the domain from where the user is coming is not
available, such as Facebook and Twitter clients or URL shortening services, we performed a series of experiments with the aim of predicting from which
referrer URL the user joined the network, \ie, if a model can predict reliably
where the user is entering our network. In general, just a few steps (\ie,
visited pages) suffice to recognize the referrer URL
correctly because the surfing behavior is very distinctive of the domain the user is
coming from. 
 
Then, using the \rgraphs, we performed several experiments using a very
large network of sites (with almost two billions of user transitions) to assess
to what extent the browsing patterns information can be generalized, if one is
only provided with information from smaller subgraphs.
First, we computed the PageRank of the subgraphs and on their step-by-step BFS
expansion, measuring the distance in terms of Kendall's $\tau$ with the
PageRank computed on the full graph. To control for the subgraph size and type,
and to study the impact of the expansion strategy on the PageRank convergence,
we used two flavors of BFS and three different sets of initial subgraphs.
We found that expanding the local graph with few nodes of largest value of
PageRank leads to a faster ($74\%$ at the first expansion step), although less accurate convergence in the long run. On the other hand, adding more nodes lead to a slower converge rate in the first steps ($65\%$). 
Therefore, in all the cases where a strong convergence with the values of
the global PageRank is not required, selecting few specific nodes is enough to
significantly improve the PageRank values of the local nodes, without having to request
and process a larger amount of data.
Finally, we considered the case of a service provider that wants to estimate the reliability of the scores of PageRank computed on its local \bgraph, with respect to the ones computed on the global graph. 
Therefore, we performed another experiment trying to predict the value of the
Kendall's $\tau$ between the local and the global PageRank, only considering
information available on the local graph. We explored six different sets of
topological and structural features of the browse graph, namely size and
connectivity, assortativity, degree, weighted degree, local PageRank and
closeness. Then we computed those features on a training set that we obtained by applying
a jackknife sampling of our subgraphs, and we ran a regression on the Kendall's
$\tau$ of the PageRank of the full subgraph and the various samplings. We found
that a random forest ensemble built on \emph{weighted degree}, outperforms all
the other in terms of mean square error. When applying the regression to the
task of predicting the $\tau$ value of the global graph with the seven subgraphs
at hand, we were able to reproduce quite well the ranking of their estimated
$\tau$ values with their actual ranking, up to a Spearman's coefficient of 0.8. \\

\noindent
\textbf{Future Work. }
We envision different routes worth being taken into consideration for future work.
One line of research we plan to investigate deals with the problem of user browsing prediction. In other words, what extent it may be possible to identify what are the most common patterns of topical behavior in the network and also, to build per-user browsing models to predict what would be the page to be visited next. 
Further, motivated by real use case scenarios,
we considered subgraphs determined by the referrer URL of user sessions;
we believe that interesting analytical results could be found, when considering
other types of subgraphs, such as networks induced by nodes that belong to the same
topical area.

\section{Acknowledgments} 
This work was partially funded by Grant TIN2009-14560-C03-01 of the Ministry of Science and Innovation of Spain, by the EU-FET grant NADINE (GA 288956) and by a Yahoo Faculty Research Engagement Program. 

\balance
{\small
\bibliographystyle{abbrv}
\bibliography{lrpaper}  
}

\end{document}

%% file: macros.tex
\usepackage[latin]{babel}
\usepackage{hyphenat}
\usepackage{rotating}
\usepackage{booktabs} 
\usepackage{subfig}
\usepackage[table,usenames,dvipsnames]{xcolor}
\usepackage{color}
\usepackage[hyphens]{url}
\usepackage[colorlinks, citecolor=blue, linkcolor=brown, urlcolor=gray]{hyperref}
\usepackage{paralist}
\usepackage[show]{chato-notes} 
\usepackage{multirow}
\usepackage{hhline}
\usepackage{float}

\usepackage{pseudocode}
\usepackage[ruled,vlined]{algorithm2e}
\newcommand{\superscript}[1]{\ensuremath{^{\textrm{#1}}}}
\def\sharedaffiliation{\end{tabular}\newline\begin{tabular}{c}}
\def\upf{\superscript{1}}
\newcommand{\ie}{\emph{i.e.}}
\newcommand{\eg}{\emph{e.g.}}
\newcommand{\etc}{\emph{etc.}}
\newcommand{\etal}{\emph{et al.}}

\newcommand{\browsegraph}{\textit{BrowseGraph}}

\newcommand{\rgraph}{\textit{ReferrerGraph}}
\newcommand{\rgraphs}{\textit{ReferrerGraphs}}
\newcommand{\bgraph}{\textit{BrowseGraph}}
\newcommand{\bgraphs}{\textit{BrowseGraphs}}

%% file: lrpaper-fp113.bbl
\begin{thebibliography}{10}
\vspace{0.4cm}

\bibitem{Agichtein:2006:IWS:1148170.1148177}
E.~Agichtein, E.~Brill, and S.~Dumais.
\newblock Improving web search ranking by incorporating user behavior
  information.
\newblock In {\em SIGIR}, pages 19--26, New York, NY, USA, 2006. ACM.

\bibitem{Andersen2007}
R.~Andersen, C.~Borgs, J.~Chayes, J.~Hopcraft, V.~S. Mirrokni, and S.-H. Teng.
\newblock Local computation of pagerank contributions.
\newblock In {\em WAW}, pages 150--165, San Diego, CA, USA, 2007.
  Springer-Verlag.

\bibitem{Bar-Yossef2008}
Z.~Bar-Yossef and L.-T. Mashiach.
\newblock Local approximation of pagerank and reverse pagerank.
\newblock In {\em CIKM}, pages 279--288, Napa Valley, California, USA, 2008.
  ACM Press.

\bibitem{Barrat2008}
A.~Barrat, M.~Barthlemy, and A.~Vespignani.
\newblock {\em Dynamical Processes on Complex Networks}.
\newblock Cambridge University Press, New York, NY, USA, 2008.

\bibitem{Bharat1998}
K.~Bharat, A.~Broder, M.~Henzinger, P.~Kumar, and S.~Venkatasubramanian.
\newblock The connectivity server: fast access to linkage information on the
  web.
\newblock In {\em WWW}, volume~30, pages 469--477, Brisbane, Australia, 4 1998.
  Elsevier Science Publ B. V.

\bibitem{BoldiWAW2004}
P.~Boldi, M.~Santini, and S.~Vigna.
\newblock Do your worst to make the best : Paradoxical effects in pagerank
  incremental computations.
\newblock In {\em WAW}, pages 168--180. Springer, 2004.

\bibitem{Breiman:2001:RF:570181.570182}
L.~Breiman.
\newblock Random forests.
\newblock {\em Machine Learning}, 45(1):5--32, oct 2001.

\bibitem{BressanWWW13}
M.~Bressan, E.~Peserico, U.~Padova, and L.~Pretto.
\newblock The power of local information in pagerank.
\newblock In {\em WWW Companion}, pages 179--180, Rio de Janeiro, Brazil, 2013.

\bibitem{Bressan2011}
M.~Bressan and L.~Pretto.
\newblock Local computation of pagerank: the ranking side.
\newblock In {\em CIKM}, pages 631--640. ACM, 2011.

\bibitem{Chen2004}
Y.-Y. Chen, Q.~Gan, and T.~Suel.
\newblock Local methods for estimating pagerank values.
\newblock In {\em CIKM}, pages 381--389, New York, NY, USA, 2004. ACM.

\bibitem{chiarandiniICWSM13}
L.~Chiarandini, P.~Grabowicz, M.~Trevisiol, and A.~Jaimes.
\newblock Leveraging browsing patterns for topic discovery and photostream
  recommendation.
\newblock In {\em ICWSM}, Cambridge, MA, USA, 2013. AAAI.

\bibitem{chiarandini12discovering}
L.~Chiarandini, M.~Trevisiol, and A.~Jaimes.
\newblock Discovering social photo navigation patterns.
\newblock In {\em ICME}, pages 31--36. IEEE, 2012.

\bibitem{Cho1998}
J.~Cho, H.~Garcia-Molina, and L.~Page.
\newblock Efficient crawling through url ordering.
\newblock In {\em WWW}, volume~30, pages 161--172, Brisbane, Australia, 4 1998.
  Elsevier Science Publ B. V.

\bibitem{DavisKDD06}
J.~V. Davis and I.~S. Dhillon.
\newblock Large scale analysis of web revisitation patterns.
\newblock In {\em KDD}, volume~08, pages 116--125, Philadelphia, PA, USA, 2006.
  ACM Press.

\bibitem{Gyongyi2004}
Z.~Gy\"ongyi, H.~Garcia-Molina, and J.~Pedersen.
\newblock Combating web spam with trustrank.
\newblock In {\em VLDB}, pages 576--587, Toronto, ON, Canada, 2004.

\bibitem{hu07demographic}
J.~Hu, H.-J. Zeng, H.~Li, C.~Niu, and Z.~Chen.
\newblock Demographic prediction based on user's browsing behavior.
\newblock In {\em WWW}, pages 151--160, New York, NY, USA, 2007. ACM.

\bibitem{Kleinberg1999}
J.~Kleinberg.
\newblock Authoritative sources in a hyperlinked environment.
\newblock {\em Journal of the ACM}, 46(5):604--632, 1999.

\bibitem{lehmann2013measuring}
J.~Lehmann, M.~Lalmas, and R.~Baeza-Yates.
\newblock Measuring inter-site engagement.
\newblock In {\em Big Data, 2013 IEEE International Conference on}, pages
  228--236. IEEE, 2014.

\bibitem{Lempel2001}
R.~Lempel and S.~Moran.
\newblock Salsa : The stochastic approach for link- structure analysis.
\newblock {\em Challenge}, 19(2):131--160, 2001.

\bibitem{liu10personalized}
J.~Liu, P.~Dolan, and E.~R. Pedersen.
\newblock Personalized news recommendation based on click behavior.
\newblock In {\em IUI}, pages 31--40, New York, NY, USA, 2010. ACM.

\bibitem{liu11user}
M.~Liu, R.~Cai, M.~Zhang, and L.~Zhang.
\newblock User browsing behavior-driven web crawling.
\newblock In {\em CIKM}, pages 87--92, New York, NY, USA, 2011. ACM.

\bibitem{Liu2008a}
Y.~Liu, B.~Gao, T.-Y. Liu, Y.~Zhang, Z.~Ma, S.~He, and H.~Li.
\newblock Browserank: letting web users vote for page importance.
\newblock {\em SIGIR}, 31:451--458, 2008.

\bibitem{Liu2009}
Y.~Liu, T.-Y. Liu, B.~Gao, Z.~Ma, and H.~Li.
\newblock A framework to compute page importance based on user behaviors.
\newblock {\em Information Retrieval}, 13(1):22--45, 6 2009.

\bibitem{Page1998}
L.~Page, S.~Brin, R.~Motwani, and T.~Winograd.
\newblock The pagerank citation ranking: Bringing order to the web.
\newblock {\em World Wide Web Internet And Web Information Systems},
  54(2):1--17, 1998.

\bibitem{Smola03atutorial}
A.~J. Smola and B.~Sch\"olkopf.
\newblock A tutorial on support vector regression.
\newblock Technical report, Statistics and Computing, 2003.

\bibitem{Strobl2008rf}
C.~Strobl, A.-L. Boulesteix, T.~Kneib, T.~Augustin, and A.~Zeileis.
\newblock Conditional variable importance for random forests.
\newblock {\em BMC Bioinformatics}, 9(1):307, 2008.

\bibitem{TrevisiolRecSys14}
M.~Trevisiol, L.~M. Aiello, R.~Schifanella, and A.~Jaimes.
\newblock Cold-start news recommendation with domain-dependent browse graph.
\newblock In {\em RecSys}, Foster City, CA, 2014. ACM.

\bibitem{trevisiol12image}
M.~Trevisiol, L.~Chiarandini, L.~M. Aiello, and A.~Jaimes.
\newblock Image ranking based on user browsing behavior.
\newblock In {\em SIGIR}, pages 445--454, New York, NY, USA, 2012. ACM.

\bibitem{Tsagkias:2012:LIM:2348283.2348330}
M.~Tsagkias and R.~Blanco.
\newblock Language intent models for inferring user browsing behavior.
\newblock In {\em SIGIR}, pages 335--344, New York, NY, USA, 2012. ACM.

\bibitem{wasserman94social}
S.~Wasserman and K.~Faust.
\newblock {\em Social Network Analysis: Methods and Applications}.
\newblock Cambridge Univ. Press, 1994.

\bibitem{WeischedelWOW2006}
B.~Weischedel and E.~K. R.~E. Huizingh.
\newblock Website optimization with web metrics: A case study.
\newblock In {\em ICEC}, pages 463--470, New York, NY, USA, 2006. ACM.

\bibitem{White07investigatingbehavioral}
R.~W. White.
\newblock Investigating behavioral variability in web search.
\newblock In {\em In Proc. WWW}, pages 21--30, 2007.

\bibitem{White:2010:ASR:1835449.1835548}
R.~W. White and J.~Huang.
\newblock Assessing the scenic route: measuring the value of search trails in
  web logs.
\newblock In {\em SIGIR}, pages 587--594, New York, USA, 2010. ACM.

\end{thebibliography}
